# Tunneling Spectroscopy of Graphene and related Reconstructions on SiC(0001)

Shu Nie and R. M. Feenstra[*]
Dept. Physics, Carnegie Mellon University, Pittsburgh, PA 15213

## Abstract

The 5×5, 6√3×6√3-R30°, and graphene-covered 6√3×6√3-R30° reconstructions of the SiC(0001) surface are studied by scanning tunneling microscopy and spectroscopy. For the 5×5 structure a rich spectrum of surface states is obtained, with one state in particular found to be localized on top of structural protrusions (adatoms) observed on the surface. Similar spectra are observed on the bare 6√3×6√3-R30° reconstruction, and in both cases the spectra display nearly zero conductivity at the Fermi-level. When graphene covers the 6√3×6√3-R30° surface the conductivity at the Fermi-level shows a marked increase, and additionally the various surface state peaks seen in the spectrum shift in energy and fall in intensity. The influence of the overlying graphene on the electronic properties of the interface is discussed, as are possible models for the interface structure.

## I. Introduction

Considerable interest has been shown over the past few years in the properties of epitaxial graphene (monolayers of graphite) on the SiC(0001) surface, owing to the novel electronic behavior of this material.[1] Graphene forms as a very carbon-rich reconstruction of the SiC(0001) surface, and a number of other, less carbon-rich, structures also exist for this surface. Early work by Owman and Mårtensson provided scanning tunneling microscopy (STM) images of exceptional clarity for many of these structures, including the 5×5 and 6√3×6√3-R30° reconstructions.[2,3,4] The latter structure, which for convenience we refer to simply as 6√3, has subsequently been studied in great detail by many workers.[5,6,7,8,9,10,11,12,13,14] The 6√3 structure initially forms on the surface in the absence of graphene, although subsequent coverage by graphene is found to largely preserve the structure. Hence, the 6√3 reconstruction determines the atomic arrangement of the graphene/SiC interface, the structure of which may be relevant in limiting the transport properties of electrons in the graphene. However, no accepted structure for the 6√3 surface exists, despite a few proposed models.[5,7,11]

   In this work we report on scanning tunneling microscopy/spectroscopy (STM/STS) measurements of 5×5, 6√3, and graphene-covered 6√3 surfaces. A number of previously unreported surface state features are observed in the tunneling spectra, and we investigate for the 5×5 surface in particular the spatial dependence of the spectra. A particular filled state feature at about 1.1 eV below the Fermi-level is found to be localized on top of structural protrusions seen on the surface, and we associate these protrusions with adatoms. A very similar spectra is found on the 6√3 surface, despite the apparently quite different STM image for that surface. When graphene covers the 6√3 surface, the spectral features are found to shift slightly in their energy and fall in intensity. The resulting spectra are in good agreement with recent results of Brar et al.[12] for bilayer graphene, and like those authors we associate the features in the spectra with the $6\sqrt{3}$ states at the graphene/SiC interface. We discuss the evolution of the spectra with graphene coverage in terms of possible indirect effects of the overlying graphene layer(s). Our STM images for the 5×5 and $6\sqrt{3}$ surface are in good agreement with prior work,[2,3] revealing adatom positions that are inconsistent with simple adatoms on a bulk-terminated SiC bilayer. Based on the adatom positions, for the 5×5 surface particularly, we suggest that a complex carbon-rich adlayer may be present below the adatoms for both surfaces.

[*] corresponding author: feenstra@cmu.edu

## II. Experimental

Our experiments are performed on a home-built room-temperature STM that has been previously described.[15] Pt-Ir probe-tips are employed, and they are thoroughly cleaned by electron-bombardment prior to use. Images are acquired with a constant current of 0.1 nA, and at sample-tip voltages specified below. Conductance spectra are measured using a lock-in amplifier with 50 mV modulation at a frequency of typically 1 kHz. The method of variable tip-sample separation was employed,[16] using a ramp of typically 0.1 nm/V and with broadening parameter of 0.5 V for computing the normalized conductance[17] (the results are only weakly dependent on these parameter values). Spectra for each of the 5×5, $6\sqrt{3}$, and graphene-covered $6\sqrt{3}$ surfaces were acquired with at least four different probe-tips; the tip-to-tip variation in the spectra is found to be considerably smaller than the change in the spectra between the former two surfaces compared to the latter. Nominally on axis 6H-SiC(0001) substrates, n-type with resistivity in the range 0.1–1 ohm-cm, were used. Prior to study, the substrates were H-etched in a hydrogen flow of 10 lpm at 1600°C for 100 s to eliminate scratches left by polishing damage on the as-received substrates.[18,19] The H-etched substrates were then loaded into a UHV chamber and annealed by resistive heating, with temperatures measured using a disappearing-filament pyrometer pyrometer and assuming an effective emissivity of 0.90.6,[20] Low-energy electron diffraction (LEED) is used to monitor the formation of the surface structures.[21] The well known $\sqrt{3} \times \sqrt{3}$-R30° pattern forms at a temperature of about 900°C (annealing time $\approx$30 min), and of particular interest is the pattern of $6\sqrt{3} \times 6\sqrt{3}$-R30° spots that form around the $\sqrt{3}$ spots at about 1200°C. At higher temperatures the intensity of the $\sqrt{3}$ spots decreases relative to the surrounding $6\sqrt{3}$ spots thus providing a sensitive measure of the surface structure,[2,3] as further described below.

## III. Results

### A. 5×5 Surface

Our observations of the 5×5 structure were obtained from samples annealed at about 1250°C, for which the $\sqrt{3}$ LEED spots are visible but weak compared to their neighboring $6\sqrt{3}$ spots. We find that STM images of these surfaces reveal roughly equal proportions of 5×5 and 6×6 reconstruction, along with a small amount of $\sqrt{3} \times \sqrt{3}$-R30° structure. Figure 1(a) displays an image of such a surface, revealing both 5×5 and $6\sqrt{3} \times 6\sqrt{3}$-R30°. The 5×5 region is seen to be composed of a series of topographic protrusions, which we refer to as multimers. Expanded views of the 5×5 structure are shown in Figs. 1(b)–(e), displaying the results with gray scale computed according to surface curvature thereby enabling a clear discrimination of the subunits within each multimer. A dimer, trimer, and tetramer are shown in panels (c)–(e), respectively, and on those images we superimpose a grid with spacing corresponding to 2× the primitive SiC surface cell. The nearest-neighbor distance between subunits in the multimers is found to be 2.2±0.1 of a 1× spacing ($a$ = 0.308 nm), or 0.68 ± 0.03 nm, and neighboring subunits are positioned relative to each other along <1$\bar{1}$00> surface directions.

The corrugation seen in constant-current images of the 5×5 protrusions has relatively little voltage dependence, as reported previously of Oman et al.[2] and also measured by us (not shown). More detailed spectroscopic information can be obtained by imaging the conductance, dI/dV, as shown in Fig. 2. Figure 2(a) shows a topographic image, acquired at −1.5 V, and a simultaneously acquired conductance image is shown in Fig. 2(b). Interpretation of the latter is possible using the full spectra acquired from the same surface region, Fig. 2(c). A spectrum acquired *on top* of a multimer is shown by the solid line, and a spectrum acquired *between* multimers is shown by the dashed line. We find spectral peaks at − 2.0 , − 1.1 , − 0.5 , and about + 0.5 V. The −1.1 V peak in particular is present on top of the multimers, but not between them. A minimum in the conductance occurs at lower voltages of −1.5 V, and it is this minimum that produces the distinct minimum in the conductance seen on top of the multimers in Fig. 2(b). A pair of spectra acquired from a different sample using a different probe tip is shown in Fig. 2(d). Some shift in the position of spectral peaks occurs here (likely arising from an unintentional voltage drop in tip or sample), as is characteristic of tip-to-tip reproducibility, but the overall spectral shapes are similar to that of Fig. 2(c).

## B. 6√3×6√3-R30° Surface

For higher annealing temperatures we find, in agreement with prior work,[2,3] that the intensity of the √3 spot in LEED decreases and the proportion of 6√3 structure on the surface increases. Figure 3 shows a typical STM image from the 6√3 surface prepared at 1350°C, for which the √3 spot in LEED has completely vanished. Several distinctive structural features are apparent in the image, as first observed by Owman and Martensson,[2,3] including trimers (labeled by "T") and six-pointed star shapes or asterisks (labeled by "A"). We find, in agreement with prior work,[2,3,9,11] that the shapes of such features do *not* vary with the graphene coverage of the 6√3, although the corrugation amplitude does decrease with the coverage (the particular surface shown in Fig. 3 is believed to be covered with a bilayer of graphene, as described below).

Figures 3(b)–(d) show expanded views of certain features, again with a superimposed 2×2 grid on the images. Two types of trimers occur on the surface, Figs. 3(b) and (c), with a slight twist differentiating them. The spacing of the protrusions within the trimers is found to be 1.9±0.1 of a 1× spacing, similar to that on the 5×5 surface, but with neighboring protrusions now aligned nearly along <11$\bar{2}$0>. An asterisk-feature is displayed in Fig. 3(d); a trimer of protrusions can also be faintly seen near the center of these features. An additional interesting feature of Fig. 3(a) is the two hexagonal shaped regions labeled "H", shown in an expanded view in Fig. 3(e). Within the center of each region is an array of corrugation maxima with √3 spacing and <1$\bar{1}$00> alignment (as expected for simple adatoms), and the two regions are surrounded by brighter protrusions.

Tunneling spectra obtained from 6√3 surfaces prepared at various temperatures are displayed in Fig. 4. For the 1250°C preparation, Figs. 4(a) and (b), we find a spectrum quite similar to that found for the 5×5 surface of Fig. 2(c) and (d). Peaks occur at −1.1, −0.5, and +0.6 V (using the smallest observed voltages from the measurements for each peak). The overall peaks positions and spectral intensities are, within tip-to-tip reproducibility, the same as for the 5×5 spectra of Fig. 2. The spectra of Fig. 4(b) were acquired at two nearby spatial locations, and show similar variability in the intensity of the −1.1 V peak as seen in Fig. 2. In contrast, for the sample prepared at 1350°C the spectrum is quite different, as shown in Figs. 4(c) and (d). The location of spectral peaks in Figs. 4(c) and (d) has shifted somewhat compared to Figs. 4(a) and (b), and the peaks have much smaller amplitude. Importantly, Figs. 4(c) and (d) shows distinctly metallic behavior with values of normalized conductance of unity at 0 V, in contrast the spectra of Figs. 4(a) and (b) which show values of the normalized conductance at 0 V of ≈0.3, far below unity.

We interpret the differences between the spectra of Fig. 4 in terms of differing numbers of graphene layers covering the basic 6√3 structure, with the results of Figs. 4(a) and (b) corresponding to zero graphene layers and those of Figs. 4(c) and (d) corresponding to at least one, and more likely two, graphene layers. As is well known, at sufficiently high temperature the 6√3×6√3-R30° structure becomes covered with one or more layers of graphene. In principle the difference between 1 or 2 graphene layers can be identified by direct STM imaging.[12] We do indeed observe the graphene lattice in our STM images at reduced voltages (not shown), but the graphene corrugation amplitude is <0.01 nm and hence we are unable to use those images to distinguish between single or bilayer graphene. An additional means of ascertaining different graphene coverage is based on the corrugation amplitudes for the 6√3×6√3-R30°.[2,3,7,9,11,12] We do observe considerable variations in this corrugation amplitude from place to place on the surface, but again we are unable to use this as an absolute determination of coverage since the amplitude also varies considerably with the sharpness of the probe tip. In any case, based on the annealing temperature and LEED results, we expect for our surfaces prepared at 1250°C that graphene will *not* be covering most of the 6√3 structure whereas at temperatures of 1350°C we *do* expect such coverage (and we consider it likely to have more than a single graphene layer on the surface).

The graphene spectra shown in Figs. 4(c) and (d) are very similar to the results of Brar et al. (their spectra shown only over –0.4 to +0.4 V) for bilayer graphene,[12] thus indicating that our surface does in fact have two layers of graphene. As shown by those authors, some small variation does occur from point to point in the spectra, and an example of this type of variation is shown in Fig. 5, again acquired from the same surface region as Fig 3. A 5×5 grid of spectra were acquired at points shown in the image, and some of those spectra from specific spatial points are displayed: from a trimer (A), from adatoms around the observed hexagonal feature (B), and from corrugation minima in the image (C). Distinct variations are seen in the spectral results, particularly near and below 0 V. The fact that the spectral features vary from point to point provides evidence that these features arise from the underlying atomic arrangement in the 6√3 structure at the graphene/SiC interface (as also concluded by Brar et al.[12]), as opposed to some band feature of the graphene itself. An average of all 25 spectra from the image of Fig. 5 is given by spectrum (c) of Fig. 4.

**IV. Discussion**

The evolution of the spectra that we observe here is, we believe, consistent with the known phenomena of graphene layers covering the bare $6\sqrt{3}$ structure. The graphene band structure is well known, with Dirac points at the K and K' point at the edge of the first Brillouin zone, and possibly with small gaps (depending on conditions) opening up around these Dirac points.[22] The $6\sqrt{3}$ states will exist at the graphene/SiC interface; these states are expected to have quite low wavevectors compared to the graphene near-zone-edge states and so we do not expect much mixing between the two types of states. Hence, the $6\sqrt{3}$ states will propagate through the graphene up to the surface, and can be detected there by STS. Compared to the bare $6\sqrt{3}$ surface we expect a reduced intensity of the $6\sqrt{3}$ spectral features due to their propagation through the graphene, and additionally the spectra for the graphene-covered surface will display the metallic conduction associated with graphene (which has recently been shown to arise from a phonon-assisted second order tunneling process[23]).

Comparing our spectra with this expectation, we focus on the highest-lying occupied state in Fig. 4, seen at about $-0.5$ V for the bare $6\sqrt{3}$ surface and at $-0.2$ V for the graphene-covered $6\sqrt{3}$ surface. The intensity of this peak does indeed show a significant decrease due to the graphene coverage, as expected, but additionally it shows a shift in energy. This shift is also apparent if we compare the first minimum in the conductance at negative voltages, seen at $-0.9$ V for Figs. 4(a) and the solid line of Fig. 4(b) compared to $-0.5$ V for Figs. 4(c) and (d). Again, we conclude that there is a shift in energy, of about 0.4 eV. Indeed, this minimum occurs in the graphene spectra at almost exactly the same place at the peak maximum for the $6\sqrt{3}$ surface, so clearly a shift occurs between the two cases.

One contribution to the observed spectral shift will come from a possible shift in Fermi-level position at the surface when graphene forms, i.e. due to charge transfer between the substrate and the graphene as has been previously discussed.[10,22] However, such charge transfer is found to be small,[22] and indeed Seyller et al. find a nearly zero (within an accuracy of $\pm 0.1$ eV) shift in band bending for the SiC bulk bands when the $6\sqrt{3}$ surface is graphitized.[8] A second possible origin for the observed shift is a redistribution in the spectral weight of the $6\sqrt{3}$ states themselves, due to the presence of the metallic graphene layers influencing the interatomic potential in some way. An additional intriguing possibility is that the $6\sqrt{3}$ surface might be a Mott-Hubbard insulator, as are the more Si-rich $\sqrt{3} \times \sqrt{3}$ and $3 \times 3$ SiC surfaces.[5,24] If the bare $6\sqrt{3}$ is indeed a Mott-Hubbard insulator, then the subsequent presence of graphene on top of this structure might influence the electron-electron correlations in such a way so as to suppress this insulating state and hence produce a *conducting* interface. Such an interface would represent a parallel conduction path for the electrons in the graphene, which could be detrimental to the achievement of high mobilities in this system.

Compared to prior measurements of tunneling spectra from the $6\sqrt{3}$ surface and/or from graphene, our results extend over a somewhat larger voltage range that emphasizes the occurrence of a range of interface states in the spectrum. Recent low-temperature STS work by Rutter et al. extending over a range of $\pm 0.5$ V revealed a band gap of width $\approx 300$ meV for the $6\sqrt{3}$ structure whereas the graphene-covered $6\sqrt{3}$ was found to be metallic.[11] Our results are consistent with theirs, given that our spectra are acquired at room temperature so that the band edges for our $6\sqrt{3}$ spectra are not so well defined. Additional low-temperature STS results for monolayer and bilayer graphene on the $6\sqrt{3}$ are provided by Brar et al.,[12] already referred to above. These workers report a $\approx$100 meV gap around 0 V in their spectra,[12] which they have recently demonstrated to arise from an inelastic phonon channel in the tunneling.[23] We do not observe this sort of feature, most likely because its effect is greatly attenuated at room temperature.

We conclude by making a few observations concerning possible structural models for the 5×5 and $6\sqrt{3}$ reconstructions. No prior proposals have been put forth for the structure of the 5×5 reconstruction, although the more Si-rich $\sqrt{3}\times\sqrt{3}$-R30° is known to consist of Si adatoms arranged on the surface,[5] with adatom coverage of 1/3 ML (ML = SiC monolayer = 12.17 atoms/nm$^2$). It should be noted that stoichiometry of the 5×5 structure is presently unknown, although the $6\sqrt{3}$ structure has recently been determined to have carbon content quite close to that of a monolayer of graphene (38.0 carbon atoms/nm$^2$).[14] The 5×5 reconstruction is found to form at temperatures between those of the $\sqrt{3}\times\sqrt{3}$ and $6\sqrt{3}$ surfaces, and thus we expect it to have stoichiometry somewhere between those two surfaces.

Considering our STM images for the 5×5 surface, it is natural to associate the observed subunits (protrusions) within each multimer with adatoms on the surface. However, the observed arrangement of the adatoms, with spacing of $(2.2\pm0.1)a$ along $<1\bar{1}00>$, provides a very important constraint on possible structures. Simple adatom arrangements would correspond to spacing of $\sqrt{3}a$ along $<1\bar{1}00>$ for $\sqrt{3}\times\sqrt{3}$-R30°, $2a$ along $<11\bar{2}0>$ for 2×2, or a mixture of those for c(2×4), all of which are inconsistent with the experiment. This situation was also ascertained by Owman and Mårtensson, who demonstrated that many of the adatom features on *both* the 5×5 and the $6\sqrt{3}$ surfaces were consistent with an underlying 2.1×2.1-R30° mesh.[2] For the $6\sqrt{3}$ surface they found long range order in this mesh, whereas for the 5×5 surface it was only found for the protrusions (adatoms) within a given multimer, with no apparent coherence between neighboring multimers.[2]

Given that the $6\sqrt{3}$ surface has close to one graphene layer of excess carbon atoms,[14] it is natural to associate those atoms with the ones responsible for the apparent 2.1×2.1-R30° mesh on the surface. There are several possibilities: the carbon monolayer might be *above* other adatoms on the surface, or *below* the adatoms, or it might have some complicated multi-level structure. The former case is what was suggested theoretically early on by Northrup and Neugebauer[5] and also recently by Rutter et al.[11] The second situation would imply an adlayer of carbon beneath a surface terminated by adatoms, somewhat analogous to the adlayer of silicon found in the model of Starke et al. for the SiC(0001)3×3 surface.[24] The third situation would imply a complex structure of the sort given by the *nanomesh* model of Chen et al.[7] Assuming that the stoichiometries of the 5×5 and $6\sqrt{3}$ structures are similar, we believe that the second possibility with an *underlying* carbon-containing adlayer is the most likely, given the observed inconsistency in the adatoms positions of both surfaces with a bulk-terminated SiC bilayer.

## V. Conclusions

Our STM images for the 5×5 reveal distinct structural protrusions with each protrusion containing a number of subunits. The observed distance between subunits is found to be inconsistent with simple adatoms on a bulk-terminated SiC bilayer, thus implying the presence of a carbon-rich adlayer forming beneath the adatoms. Tunneling spectra of the $5 \times 5$ and $6\sqrt{3} \times 6\sqrt{3}$-R30° surfaces are found to be quite similar, suggesting that the underlying structures of these two surfaces are similar. Spectra for the graphene-covered $6\sqrt{3}$ surface show a considerable evolution from those of the bare $6\sqrt{3}$ surface, with a reduction in intensity of the features along with a shift in their energies. The observed small shift is believed to be due to the influence of the graphene in modifying the interatomic potential of the $6\sqrt{3}$ atoms, or possibly by producing a change in the electron-electron correlations of this layer.

**Acknowledgements**
Discussions with V. W. Brar, G. Gu, H. Hibino, Z. Jiang, and J. E. Northrup are gratefully acknowledged. This work was supported by the National Science Foundation (grant DMR-0503748).

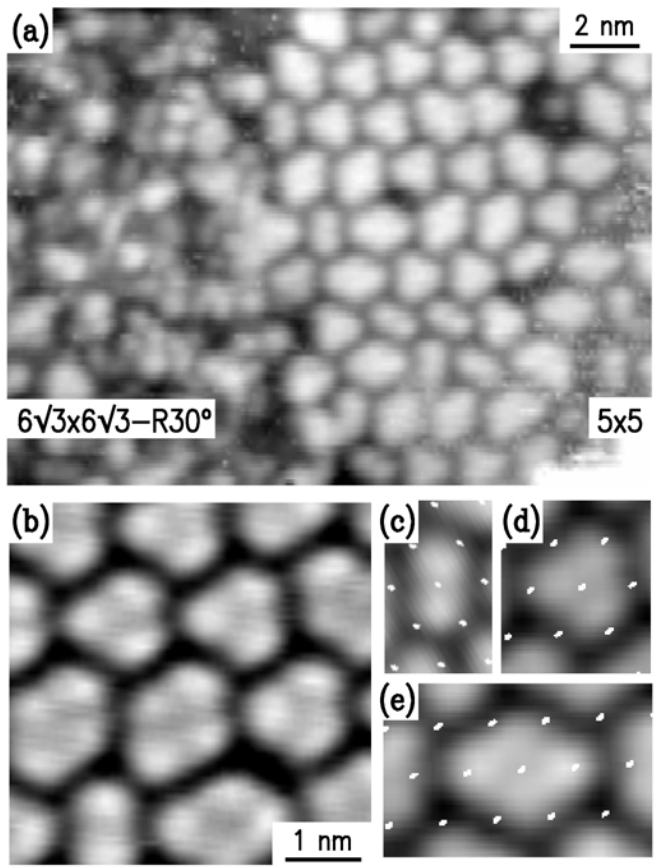

FIG 1. (a) STM image of SiC(0001) surface revealing regions of 5×5 and 6√3×6√3-R30°, structure, as indicated. The image was acquired with a sample voltage of − 2 V, and is displayed with gray scale given by surface height over a range of 0.28 nm. (b)–(e) Close-up views of the surface, displayed with grey given by surface curvature. In (c)–(e) an array of white dots is superimposed on the images, with spacing of the dots being 2×2 of the primitive SiC surface lattice (spacing between dots is 0.62 nm).

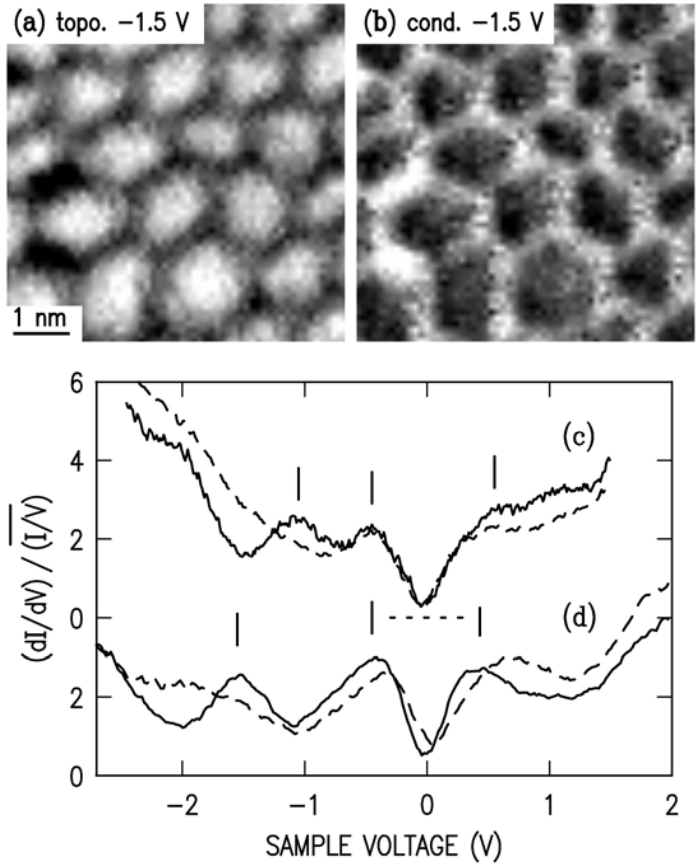

FIG 2. (a) STM topographic image and (b) conductance image, of the 5×5 surface acquired at a constant current of 0.1 nA and sample voltage of –1.5 V. (c) and (d) Tunneling spectra acquired at a topographic maximum (solid line) and a topographic minimum (dashed line). The pairs of spectra were acquired on different samples with different probe tips. Vertical lines mark peak positions, and the zero level for the upper spectra is shown by the dotted line.

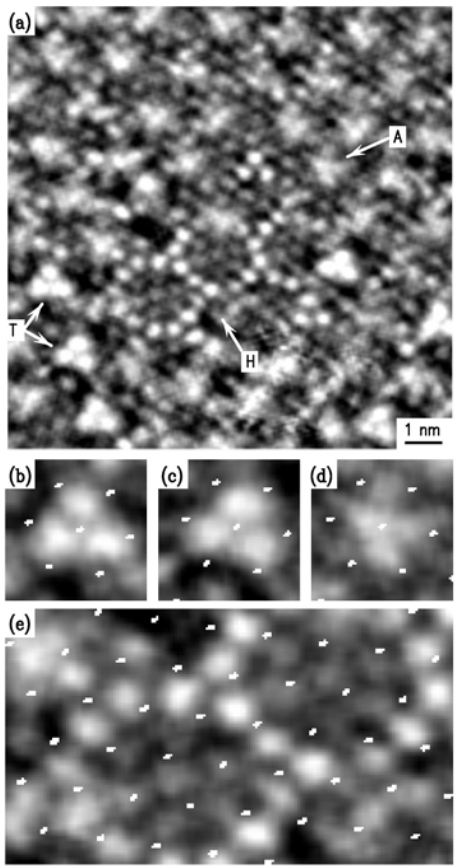

FIG 3. (a) STM image of graphene-covered 6√3×6√3 surface, acquired at −2.5 V and displayed with gray scale range of 0.10 nm. Distinctive features are indicated: clusters of three topographic maxima (T), asterisk shaped features (A), and two rings of hexagonally arranged topographic maxima (H). (b)–(e) Expanded views of trimers, asterisk feature, and hexagonally arranged corrugation maxima, with superimposed 2×2 array of dots (spacing between dots is 0.62 nm).

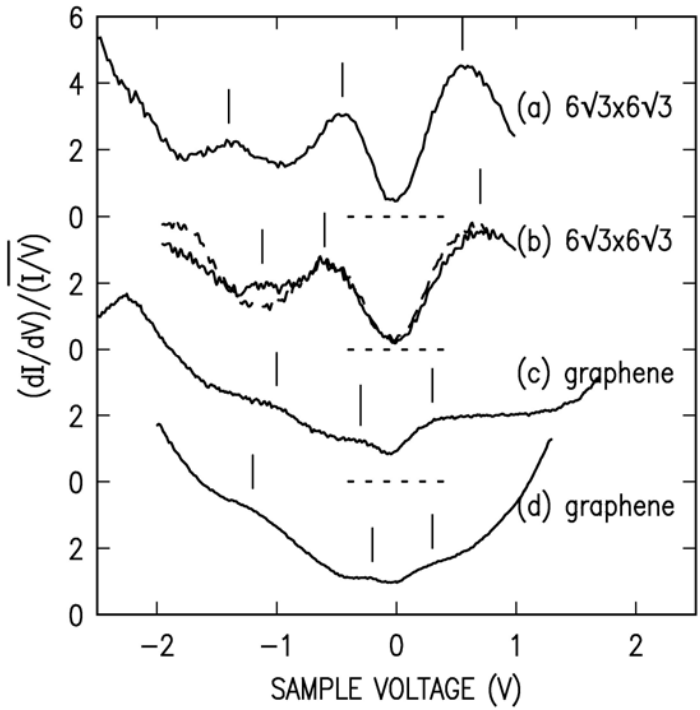

FIG 4. Tunneling spectra from surfaces prepared at (a) 1250°C, (b) 1250°C, (c) 1350°C, and (d) 1350°C. All spectra [or pair of spectra, for (b)] were acquired on different samples with different probe tips. Vertical lines mark peak positions, and the zero level for the upper spectra is shown by the dotted line.

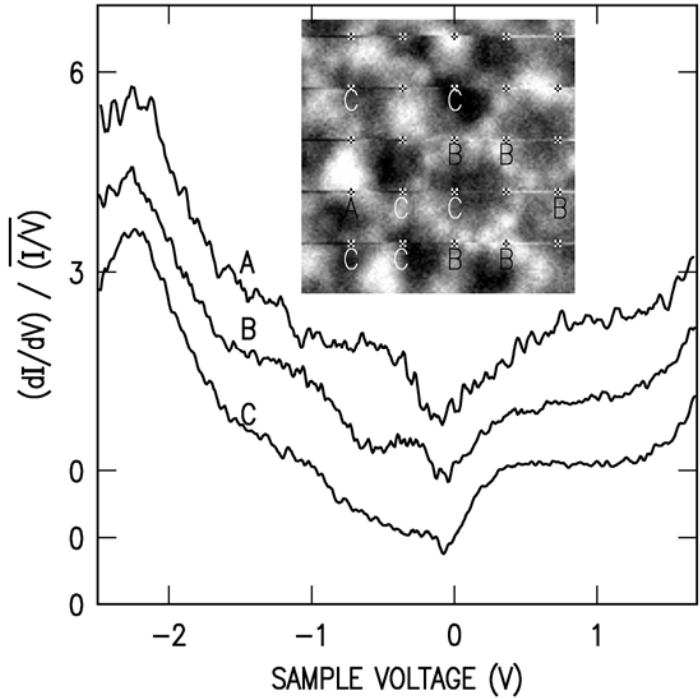

FIG 5. Spatially resolved spectra acquired from a graphene-covered $6\sqrt{3}\times6\sqrt{3}$ surface. Each spectrum A–C is an average of individual spectra acquired from the points indicated in the image (image size 6.5×6.5 nm$^2$, acquired at sample voltage of $-2.5$ V and displayed with gray scale range of 0.08 nm).

we estimate a random error of ±50°C to account for variation in temperature over the sample.